\pdfoutput=1
\documentclass[11pt]{article}

\usepackage[preprint]{acl}

\usepackage{times}
\usepackage{latexsym}

\usepackage[T1]{fontenc}
\usepackage[utf8]{inputenc}

\usepackage{microtype}

\usepackage{inconsolata}

\usepackage{CJKutf8}
\usepackage{graphicx}
\usepackage{booktabs}
\usepackage{algorithm}
\usepackage[switch]{lineno}
\usepackage{algpseudocode}
\usepackage{tcolorbox}
\usepackage{amsmath}
\usepackage{colortbl}
\usepackage{xcolor}
\definecolor{grey}{rgb}{0.3,0.3,0.3}% colors

\title{AutoMIR: Effective Zero-Shot Medical Information Retrieval without Relevance Labels}

\author{
Lei Li$^1$
\and
Xiangxu Zhang$^1$\and
Xiao Zhou$^1$\thanks{Xiao Zhou and Zheng Liu are corresponding authors.}\and
Zheng Liu$^2$\footnotemark[1]\\
$^1$Gaoling School of Artificial Intelligence, Renmin University of China\\
$^2$Beijing Academy of Artificial
Intelligence\\
\{leil, xansar, xiaozhou\}@ruc.edu.cn,
zhengliu1026@gmail.com
}

\begin{document}
\maketitle
\begin{abstract}
Medical information retrieval (MIR) is essential for retrieving relevant medical knowledge from diverse sources, including electronic health records, scientific literature, and medical databases. However, achieving effective zero-shot dense retrieval in the medical domain poses substantial challenges due to the lack of relevance-labeled data. In this paper, we introduce a novel approach called \textbf{S}elf-\textbf{L}earning \textbf{Hy}pothetical \textbf{D}ocument \textbf{E}mbeddings (\textbf{SL-HyDE}) to tackle this issue. SL-HyDE leverages large language models (LLMs) as generators to generate hypothetical documents based on a given query. These generated documents encapsulate key medical context, guiding a dense retriever in identifying the most relevant documents. The self-learning framework progressively refines both pseudo-document generation and retrieval, utilizing unlabeled medical corpora without requiring any relevance-labeled data. Additionally, we present the Chinese Medical Information Retrieval Benchmark (CMIRB), a comprehensive evaluation framework grounded in real-world medical scenarios, encompassing five tasks and ten datasets. By benchmarking ten models on CMIRB, we establish a rigorous standard for evaluating medical information retrieval systems. Experimental results demonstrate that SL-HyDE significantly surpasses HyDE in retrieval accuracy while showcasing strong generalization and scalability across various LLM and retriever configurations. Our code and data are publicly available at: \url{https://github.com/ll0ruc/AutoMIR}
\end{abstract}

\section{Introduction}
Medical information retrieval (MIR)~\cite{luo2008medsearch,goeuriot2016medical} focuses on retrieving relevant medical information from sources like electronic health records, scientific papers, and medical knowledge databases, based on specific medical queries. Its applications are wide-ranging, supporting doctors in clinical decision-making~\cite{sivarajkumar2024clinical}, assisting patients in finding health information~\cite{mcgowan2009electronic}, and aiding researchers in accessing relevant studies~\cite{zheng2015key}. 
 
Dense retrievers~\cite{karpukhin2020dense,xu2024bmretriever} have shown strong performance with large labeled datasets in information retrieval (IR). Several studies~\cite{xiong2020approximate,li2023towards,xiao2024c} have successfully employed contrastive learning to develop general-purpose text embedding models, achieving promising results in zero-resource retrieval scenarios. They leverage large-scale weakly supervised data through web crawling, or high-quality text pairs derived from data mining or manual annotation. However, the availability of such large-scale datasets cannot always be assumed, particularly in non-English languages or specialized domains.

Recently, large language models (LLMs) have demonstrated exceptional performance in zero-resource retrieval scenarios~\cite{wang2023query2doc,shen2023large,mao2024rafe}, primarily due to their extensive knowledge and robust text generation capabilities. This makes them particularly effective in situations where labeled data is scarce or unavailable. One such approach, HyDE~\cite{gao2022precise}, employs zero-shot prompts to guide an instruction-following language model to generate hypothetical documents, effectively narrowing the semantic gap between the query and the target document. Similarly, Query2doc\cite{wang2023query2doc} uses few-shot prompting of LLMs to generate pseudo-documents, which are then used to expand the original query. However, applying these methods to medical information retrieval presents three critical challenges: (1) \textbf{LLMs lack the specialized medical knowledge necessary to generate highly relevant hypothetical documents}. Although LLMs are trained on vast datasets drawn from a wide array of general-purpose sources, they are often insufficiently equipped with domain-specific knowledge, particularly in fields like medicine. (2) \textbf{General text embedding models are inadequate for representing medical queries and documents effectively}. These versatile retrievers are typically designed for multi-domain and multi-task settings, failing to capture the nuanced and knowledge-intensive nature of the medical domain. (3) \textbf{The medical domain suffers from a scarcity of high-quality, relevance-labeled datasets}. The scarcity of labeled data significantly increases the cost of training and fine-tuning these models to achieve high performance.

To address these issues, we propose \textbf{S}elf-\textbf{L}earning \textbf{Hy}pothetical \textbf{D}ocument \textbf{E}mbedding (\textbf{SL-HyDE}), an effective fully zero-shot dense retrieval system requiring no relevance-labeled data for medical information retrieval. During the inference phase, SL-HyDE first employs an LLM as the generator to produce a relevant hypothetical document in response to a medical query. A retrieval model is then employed to pinpoint the most relevant target document from the candidates based on the generated hypothetical document. In the training phase, we design a self-learning mechanism that enhances the retrieval performance of SL-HyDE without the need for labeled data. Specifically, this mechanism leverages the retrieval model's ranking capabilities to select high-relevance hypothetical documents that align with the output of the generator (LLM), simultaneously injecting medical knowledge into the LLM. In turn, the generator’s ability to produce high-quality hypothetical documents provides pseudo-labeled data for the training of retrieval model, enabling it to efficiently encode medical texts. This interactive and complementary approach generates supervisory signals that enhance both the generation and retrieval capabilities of the system. Notably, SL-HyDE begins with unlabeled medical corpora and completes the training process through a self-learning mechanism, thereby circumventing the heavy reliance on labeled data typically required for training both large language models and text embedding models.

To evaluate SL-HyDE's performance in Chinese medical information retrieval, we develop a valuable \textbf{C}hinese \textbf{M}edical \textbf{I}nformation \textbf{R}etrieval \textbf{B}enchmark (\textbf{CMIRB}). CMIRB is constructed from real-world medical scenarios, including online consultations, medical examinations, and literature retrieval. It comprises five tasks and ten datasets, marking the first comprehensive and authentic evaluation benchmark for Chinese medical information retrieval. This benchmark is poised to accelerate advancements toward more robust and generalizable MIR systems in the future.

Through extensive experimentation on the CMIRB benchmark, we find that our proposed method significantly enhances retrieval performance. We validate SL-HyDE across various configurations involving three large language models as generators and three embedding models as retrievers. Notably, SL-HyDE surpasses the HyDE (Qwen2 as generator + BGE as retriever) combination by an average of 4.9\% in NDCG@10 across ten datasets, and it shows a 7.2\% improvement compared to using BGE alone for retrieval. These outcomes underscore the effectiveness and versatility of SL-HyDE. In summary, our contributions are as follows:

\begin{itemize}
\item We propose Self-Learning Hypothetical Document Embeddings for zero-shot medical information retrieval, eliminating the need for relevance-labeled data.

\item We are the first to develop a comprehensive Chinese Medical Information Retrieval Benchmark and evaluate the performance of various text embedding models on it.

\item SL-HyDE enhances retrieval accuracy across five tasks and demonstrates generalizability and scalability with different combinations of generators and retrievers.
\end{itemize}

\section{Related Work}
\subsection{Dense Retrieval}
Recent advancements in deep learning and natural language processing have driven improvements in information retrieval. Contriever~\cite{izacard2021unsupervised} leverages unsupervised contrastive learning for dense retrieval. PEG~\cite{wu2023towards} and BGE~\cite{xiao2024c} enhance Chinese general embeddings through training on large-scale text pairs. These works demonstrate the impact of well-structured training strategies on effective retrieval across multiple domains. Beyond embedding-based techniques, large language models have demonstrated exceptional performance in zero-resource retrieval scenarios. GAR~\cite{mao-etal-2021-generation} enriches query semantics with generated content. HyDE~\cite{gao2022precise} generates hypothetical documents for the retriever, effectively narrowing the semantic gap between the query and the target document. Query2doc~\cite{wang2023query2doc} utilizes few-shot prompts to expand queries, boosting both sparse and dense retrieval. However, retrieval through hypothetical documents generated by LLMs often yields suboptimal results when domain-specific knowledge is insufficient. To address these challenges, we propose a self-learning framework that jointly optimizes the generator and retriever without any relevance labels, thereby enhancing retrieval performance.

\subsection{Information Retrieval Benchmark}
To better guide the development of retrieval models, researchers have developed various datasets and benchmarks. For instance, DuReader~\cite{he-etal-2018-dureader}, a large-scale Chinese reading comprehension dataset, significantly advances text understanding and information retrieval research.  BEIR\cite{thakur2021beir}, a zero-shot retrieval evaluation benchmark, covers diverse retrieval tasks and offers a unified evaluation platform. MTEB~\cite{muennighoff-etal-2023-mteb} establishes a framework for evaluating multilingual text embeddings. More recently, C-MTEB~\cite{xiao2024c} specifically addresses Chinese text embedding evaluations. However, these benchmarks are designed for general domains, limiting their utility for specific domains such as medical retrieval. Existing medical benchmarks like CMB~\cite{wang2024cmb} and CMExam~\cite{liu2024benchmarking} focus primarily on medical QA and clinical reasoning, which are not suitable for medical retrieval evaluation. To bridge this gap, we develop the first comprehensive and realistic evaluation benchmark based on real-world medical scenarios for Chinese medical information retrieval tasks.

\section{Methodology}

\subsection{Preliminary}
Zero-shot document retrieval is a crucial component of the search systems. Given a user query $q$ and a document set $D = \{d_1, ..., d_n\}$ where $n$ represents the number of document candidates, the goal of a retrieval model ($\mathcal{M}_r$) is to fetch documents that align with the user's genuine search intent for the current query $q$. These models map an input query $q$ and a document $d$ into a pair of vectors $⟨v_q, v_d⟩$, using their inner product as a similarity function $s(q,d)$:
\begin{equation}
\label{eq:s_qd}
  s(q,d) = <\mathcal{M}_r(q), \mathcal{M}_r(d)>.
\end{equation}

The retrieval models then identify the top-k documents, denoted as $D_{topk}$, which have the highest similarity scores when compared to the query $q$. 

Large language models have achieved remarkable success in text generation across various natural language processing tasks, including question answering~\cite{liu2021makes} and text generation~\cite{dathathri2019plug}. Recently, there has been a growing interest in utilizing these models to generate relevant documents based on queries, thereby improving retrieval accuracy. Hypothetical Document Embeddings (HyDE)~\cite{gao2022precise} decompose dense retrieval into two tasks: a generative task executed by an instruction-following language model and a document-document similarity task executed by a retrieval model.

% Earlier studies~\cite{karpukhin2020dense,ding2020rocketqa} utilized dual-tower architectures to encode queries and documents separately. However, more recent research~\cite{izacard2021unsupervised,xiao2024c} indicates that using the same architecture for both query and document encoding provides greater robustness in low-resource information retrieval tasks. In this work, we utilize the $\mathcal{M}_r$ model to serve as both the query encoder and the document encoder.

\begin{figure}
\includegraphics[width=\linewidth]{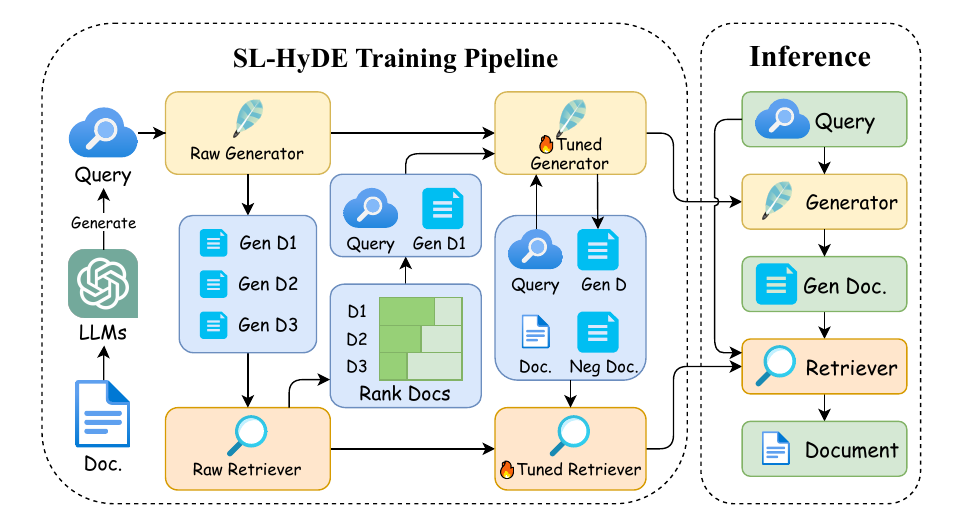}
  \caption{Training and inference pipeline of SL-HyDE.}
  \label{fig: Model}
\end{figure}

\subsection{Overview}
Applying HyDE to the medical domain presents two primary challenges: (1) LLMs often lack specialized medical domain knowledge, and (2) retrievers may struggle to effectively encode medical texts due to inadequate training on medical corpora. These challenges hinder the successful implementation of HyDE technology in the medical field, making it difficult to achieve significant performance improvements in retrieval tasks. A common strategy to supplement medical domain knowledge involves fine-tuning with labeled medical data~\cite{zhang2023huatuogpt, wang2024apollo, xu2024bmretriever}. However, these approaches rely on high-quality, manually constructed data to adapt general models to the medical domain. Unfortunately, obtaining such high-quality labeled data in practice is particularly challenging, making the training of a medical LLM highly costly.

In this paper, we introduce a self-learning hypothetical document embedding mechanism designed to leverage the potential of unlabeled medical corpora. The labels are entirely generated by the generator and retriever in SL-HyDE, eliminating the need for external labeled data collection. Figure~\ref{fig: Model} presents the overall framework. 

\subsection{SL-HyDE Training}
% Applying HyDE to the medical domain presents two primary challenges: LLMs often lack specialized medical domain knowledge, and retrievers may struggle to effectively encode medical texts due to inadequate training on medical corpora. These challenges hinder the successful implementation of HyDE technology in medical retrieval tasks. A common strategy to supplement medical domain knowledge involves fine-tuning with labeled medical data~\cite{zhang2023huatuogpt,wang2024apollo}. These approaches rely on high-quality, manually constructed data to adapt general models to the medical domain. Unfortunately, obtaining such labeled data in practice is particularly challenging, making the training of a medical LLM and text embedding model highly costly.
\noindent \textbf{Self-Learning Generator.}
An unlabeled medical corpus, such as Huatuo26M~\cite{li2023huatuo}, serves as the foundational resource for domain-specific content. To construct queries, we employ a robust offline LLM, Qwen2.5-32B-Instruct~\cite{qwen2.5}, leveraging in-context learning~\cite{brown2020language}. With a well-designed prompt, the model effectively generates medically grounded and context-aware queries:
\begin{equation}
\label{eq:}
  q = \mathrm{LLM}(d, \mathrm{prompt}).
\end{equation}

To facilitate retrieval, the raw generator creates a hypothetical document that distills the relevant information from the true target document. Concretely, we provide both the query and the corresponding target document as input to the generator, along with a carefully designed prompt to guide the generation of the pseudo-document.
\begin{equation}
\label{eq:}
  d' = \mathcal{M}_g(q, d, \mathrm{prompt}).
\end{equation}
Notably, we intentionally avoid using the true target document as the output label because the generator's primary role is to craft a hypothetical document that aids the retriever in locating it. Expecting the generator to replicate the exact target document itself would be overly demanding and unrealistic. 

Given that not all hypothetical documents generated by the generator are equally effective for retrieval, we leverage the retriever $\mathcal{M}_r$ to select the most optimal one. Specifically, the generator $\mathcal{M}_g$ creates $L$ hypothetical documents for a given query. Each hypothetical document $d'_i$ is used to retrieve documents from the corpus, and we record the rank $r_i$ of the true target document $d$. The pseudo-document with the highest retrieval quality (the lowest $r_i$) is selected:
\begin{equation}
\label{eq:}
  r_i = \mathrm{rank}(d, \mathrm{sort}(s(d'_i, D)), i=1,...,L,
\end{equation}
% \begin{equation}
% \label{eq:}
%   d^* = d'_{\mathrm{arg \ min^K_{i=1}} \ r_i}
% \end{equation}
\begin{equation}
\label{eq:}
    i^* = \mathrm{arg \ min^L_{i=1}} \ r_i, \
  d^* = d'_{i^*}.
\end{equation}

This process yields a collection of question-answer pairs in the form of $(q, d^*)$, functions as the question and the generated document as the corresponding answer. The generator is subsequently trained via supervised fine-tuning on the resulting dataset $D_{llm} = \{(q, d^*)|q \in Q\}$. The standard supervised fine-tuning (SFT) loss is computed as:
\begin{equation}
\label{eq:}
  \mathcal{L}_{\mathrm{slg}} = -\sum\nolimits_{q \in Q} \sum\nolimits_t \mathrm{log} \ \mathcal{M}_g(d'_t|d'_{\textless t}, q).
\end{equation}

Interestingly, the self-learning generator is trained without relying on supervision signals from labeled medical data. Instead, it is based on unlabeled corpora and employs the generator's text generation alongside the retriever's ranking function to construct high-quality question-answer pairs tailored for hypothetical document generation.

\noindent \textbf{Self-Learning Retriever.}
Given a passage $d$ from the corpus $D$ and its corresponding query $q$, the pair 
$(q, d)$ naturally forms the labeled query-document data required for retriever fine-tuning. However, since SL-HyDE retrieves the target document by encoding both the query and a generated hypothetical document when inference, we explore a triplet $(q, d'; d)$ as the labeled data for retriever training. This approach effectively aligns the training data format with that of the inference stage, thereby enhancing consistency and bridging the gap between training and deployment.

To achieve this, we utilize the fine-tuned generator $\mathcal{M}_g^t$ from the previous stage to generate hypothetical documents for all queries, constructing a labeled fine-tuning dataset $D_{emb} = \{(q, d'; d)|q \in Q\}$. Following previous research~\cite{li2023towards,xiao2024c}, we further increase the training data complexity through hard negative mining. Specifically, a retriever is used to identify difficult negative samples from the original corpus $D$ through an ANN-based sampling strategy~\cite{xiong2020approximate}, resulting in a hard negative dataset:
\begin{equation}
\label{eq:}
  D^- = \mathrm{ANN} (\mathcal{M}_r(q,d'), \mathcal{M}_r(D)).
\end{equation}

In addition to the negatives mined from the corpus, we also incorporate in-batch negatives. Contrastive learning loss is then applied for the supervised fine-tuning of the retriever, with the objective function formulated as follows: 
\begin{equation}
\label{eq:}
  \mathcal{L}_{\mathrm{slr}} = \mathrm{min.} \sum_{(q,d)} -\mathrm{log} \ \frac{e^{s(q,d)/\tau}}{e^{s(q,d)/\tau} + \sum\nolimits_{B \cup D^-}e^{s(q,d^-)/\tau}},
\end{equation}
where $\tau$ is the temperature coefficient, and $B$ represents the negative samples in a batch. The score $s(q, d)$ incorporates the generated document, as described in Equation~\ref{eq:s_qd}.

At this stage, we can obtain a retriever equipped with medical domain knowledge that is coherently adapted to the characteristics of retrieval queries, incorporating hypothetical documents. In SL-HyDE, the generator and retriever are trained separately in a sequential manner, allowing each component to be optimized with the most appropriate supervision signal available at its respective training phase.

\subsection{SL-HyDE Inference}
As illustrated in Figure~\ref{fig: Model}, the inference stage of SL-HyDE introduces a hypothesis generation step prior to conventional retrieval. Specifically, the input query $q$ is first rewritten by a fine-tuned generator $\mathcal{M}_g^t$ to produce a pseudo-document $d'$, as defined by the following equation:
\begin{equation}
\label{eq:}
  d' = \mathcal{M}_g^t(q, \mathrm{prompt}).
\end{equation}
The prompt is a manually designed instruction tailored to the requirements of each task. Detailed formulations of the prompts used in our experiments are provided in Appendix~\ref{app:evaluation}.

To better fuse the documents, we sample $N$ documents from the hypothetical documents. Subsequently, a tuned retriever $\mathcal{M}_r^t$ is used to encode these documents into an embedding vector $v_q$: 
\begin{equation}
  v_q = \frac{1}{N+1} [\sum_{k=1}^N \mathcal{M}_r^t(d'_k) + \mathcal{M}_r^t(q)].
\end{equation}

Then, the inner product is computed between $v_q$ and all document vectors:
\begin{equation}
  \label{eq: sqd}
  s(q,d) = <v_q, \mathcal{M}_r^t(d)>, \forall d \in D.
\end{equation}

This vector identifies a neighborhood in the corpus embedding space, from which similar real documents are retrieved based on vector similarity. 

\section{CMIRB Benchmark}
\subsection{Overview}
% The CMIRB benchmark is a specialized multi-task dataset designed specifically for medical information retrieval. Our collection and construction methodology is guided by the following principles: (i) Domain Specificity: The benchmark focuses exclusively on medical tasks, providing a targeted evaluation setting. (ii) Task Diversity: It includes a variety of query types and document formats to reflect real-world medical scenarios. (iii) Task Difficulty: The tasks are crafted to challenge existing models, thereby pushing the boundaries of medical retrieval research. Figure~\ref{fig: bench} provides an verview of the tasks and datasets in CMIRB. 

\begin{figure}[tb]
\centering
  \includegraphics[width=\linewidth]{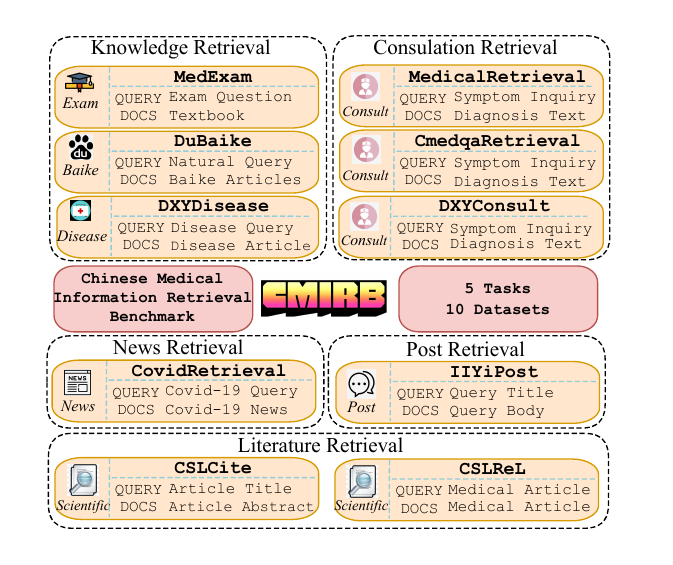}
  \caption{An overview of CMIRB.}
  \label{fig: bench}
\end{figure}

The CMIRB benchmark is a specialized multi-task dataset designed specifically for medical information retrieval. As shown in Figure~\ref{fig: bench}, it comprises five different tasks. Medical knowledge retrieval task: Retrieve relevant medical knowledge snippets from textbooks or encyclopedias based on a given medical query. Medical consultation retrieval task: Extract relevant doctor's responses to online medical consultation questions posed by patients. Medical news retrieval task: Focus on retrieving news articles that address queries related to COVID-19. Medical post retrieval task: Retrieve the content of a forum post corresponding to its title. Medical literature retrieval task: Retrieve abstracts of cited references based on a medical title or find a similar paper based on the given medical paper. 

\subsection{Data Construction}
The CMIRB benchmark integrates 10 datasets, including several existing resources: \textbf{MedicalRetrieval}~\cite{long2022multi}, \textbf{CmedqaRetrieval}~\cite{qiu2022dureader_retrieval}, and \textbf{CovidRetrieval}~\cite{qiu2022dureader_retrieval}, covering patient-doctor consultations and COVID-19-related news retrieval.

In addition, we construct several datasets by combining existing query resources with curated medical corpora. \textbf{MedExam} pairs questions with textbook passages from MedQA~\cite{jin2021disease}. \textbf{DuBaike} uses queries from DuReader\cite{he2017dureader} and documents collected from Baidu Baike pages\footnote{\href{https://baike.baidu.com/}{https://baike.baidu.com/}}. We also curate two datasets from the medical website DingXiangYuan\footnote{\href{https://dxy.com/}{https://dxy.com/}}. \textbf{DXYDisease} focuses on structured disease-related Q\&A, while \textbf{DXYConsult} captures richer patient-doctor dialogues that include symptom descriptions, medication history, and diagnostic queries. We curate \textbf{IIYiPost} by crawling posts from the IIYi forum\footnote{\href{https://bbs.iiyi.com/}{https://bbs.iiyi.com/}}.

Finally, CSLCite and CSLRel are constructed based on the CSL dataset~\cite{li2022csl}, targeting different literature retrieval scenarios. \textbf{CSLCite} uses journal titles as queries and their cited references from WanFangMedical\footnote{\href{https://med.wanfangdata.com.cn/}{https://med.wanfangdata.com.cn/}} as documents, while \textbf{CSLRel} pairs each paper with the most relevant similar paper recommended by the platform.

To ensure quality, we apply ChatGPT to exclude non-medical content and low-quality query-document pairs. Additional query-document matching is performed for MedExam and DuBaike to ensure content relevance. Full details are provided in the Appendix~\ref{sec: appendix-data}. Table~\ref{tab:dataset} summarizes dataset statistics, revealing broad variability in query and document length, ranging from short titles to long passages, ensuring the benchmark’s diversity and practical relevance.

\begin{table}
  \resizebox{\linewidth}{!}{
  \begin{tabular}{l|l|rr|rr}
  \toprule
     ~ & ~ &\multicolumn{2}{c|}{\#Samples} &\multicolumn{2}{c}{ Avg. Word Lengths} \\
     \hline
     Task & Dataset &\#Query  &\#Document  &Query  &Document  \\
   \hline
   Medical  &MedExam &697 &27,871 &96.9 &493.7 \\
   Knowledge  &DuBaike &318 &56,441 &7.6 &403.3 \\
 Retrieval &DXYDisease &1,255 &54,021 &24.3  &191.1 \\
 \hline
  Medical  &MedicalRet. &1,000 &100,999 &17.9 &122.0 \\
   Consultation  &CmedqaRet. &3,999 &100,001 &48.4 &307.7 \\
 Retrieval &DXYConsult &943 &12,577 &170.4 &370.1 \\
 \hline
  News Ret.  &CovidRet. &949 &100,001 &25.9 &332.4 \\
 \hline
  Post Ret.  &IIYiPost &789 &27,570 &15.9 &150.1 \\  
  \hline
  Literature  &CSLCite &573 &36,703 &21.9 &269.6 \\  
  Retrieval  &CSLRel &439 &36,758 &281.8 &292.2 \\  
\bottomrule
\end{tabular}
}
\caption{ Statistics of datasets in CMIRB.}
  \label{tab:dataset}
\end{table}

\begin{table*}
  \resizebox{\linewidth}{!}{
  \begin{tabular}{l|ccc|ccc|c|c|cc|l}
  \toprule
     Task   &\multicolumn{3}{c|}{Knowledge Retrieval} &\multicolumn{3}{|c|}{Consulation Retrieval} &News & Post & \multicolumn{2}{|c|}{Literature Retrieval} & \\
     \hline
     Dataset &MedExam &DuBaike &DXYDis. &Medical &Cmedqa &DXYCon. &Covid &IIYiPost & CSLCite &CSLRel &Average \\
     \hline
   Text2Vec(large) &41.39 & 21.13 & 41.52 & 30.93 & 15.53 & 21.92 & 60.48 & 29.47 & 20.21 & 23.01 & 30.56 \\
    mContriever &51.50  & 22.25 & 44.34 & 38.50  & 22.71 & 20.04 & 56.01 & 28.11 & 34.59 & 33.95 & 35.20 \\
    BM25 &31.95 & 17.89 & 40.12 & 29.33 & 6.83  & 17.78 & 78.90  & 66.95 & 33.74 & 29.97 & 35.35 \\
    OpenAI-Ada-002 &53.48 & 43.12 & 58.72 & 37.92 & 22.36 & 27.69 & 57.21 & 48.60  & 32.97 & 43.40  & 42.55 \\
     M3E(large)&33.29 & 46.48 & 62.57 & 48.66 & 30.73 & 41.05 & 61.33 & 45.03 & 35.79 & 47.54 & 45.25 \\
    mE5(large)&53.96 &53.27 & 72.10  & 51.47 & 28.67 & 41.35 & 75.54 & 63.86 & 42.65 & 37.94 & 52.08 \\
   piccolo(large) &43.11 & 45.91 & 70.69 & 59.04 & 41.99 & 47.35 & 85.04 & 65.89 & 44.31 & 44.21 & 54.75 \\
   GTE(large) &41.22 & 42.66 & 70.59 & 62.88 & 43.15 & 46.30  &88.41 & 63.02 & 46.40  & 49.32 & 55.40 \\
   BGE(large) &58.61 & 44.26 & 71.71 & 59.60  & 42.57 & 47.73 & 73.33 & 67.13 & 43.27 & 45.79 & 55.40 \\
   PEG(large) &52.78 & 51.68 & 77.38 & 60.96 & 44.42 & 49.30  & 82.56 & 70.38 & 44.74 & 40.38 & 57.46 \\
   \hline
   BGE(large) &58.61 & 44.26 & 71.71 & 59.60  & 42.57 & 47.73 & 73.33 & 67.13 & 43.27 & 45.79 & 55.40 \\
    HyDE &64.39 & 52.73 & 73.98 & 57.27 & 38.52 & 47.11 & 74.32 & 73.07 & 46.16 & 38.68 & 56.62 \\
    SL-HyDE &71.49$^*$ & 60.96$^*$ & 75.34$^*$ & 58.58$^*$ & 39.07$^*$ & 50.13$^*$ & 76.95$^*$ & 73.81$^*$ & 46.78$^*$ & 40.71$^*$ & 59.38$^*$ \\
    Improve. &$\uparrow$ \textbf{11.03\%} & $\uparrow$ \textbf{15.61\%} & $\uparrow$ \textbf{1.84\%} & $\uparrow$ \textbf{2.29\%} & $\uparrow$ \textbf{1.43\%} & $\uparrow$ \textbf{6.41\%} & $\uparrow$ \textbf{3.54\%} & $\uparrow$ \textbf{1.01\%} & $\uparrow$ \textbf{1.34\%} & $\uparrow$ \textbf{5.25\%} & $\uparrow$ \textbf{4.87\%} \\
\bottomrule
\end{tabular}
}
 \caption{Performance of various Retrieval models on nDCG@10. The first part shows ten base retrieval models, and the second shows retrieval models enhanced by hypothetical documents. $^*$ denotes the result outperforms baseline models in
 t-test at $p<0.05$ level.}
 \label{tab:WHOLE}
\end{table*}

% To ensure dataset quality, we use ChatGPT to exclude non-medical data and remove low-quality query-document pairs. For the MedExam and DuBaike datasets, beyond the initial filtering, we implement a query-document matching process to find the most similar positive document for each query. More details about the data collection sources, filtering process, and sample information can be found in the \textbf{supplementary materials}. After processing, we compile a total of 10 datasets, with statistical details presented in Table~\ref{tab:dataset}. Statistics reveal that query lengths vary significantly, ranging from a single title to a full article. Similarly, document lengths span from brief physician responses to extensive medical knowledge texts. This diversity ensures that our benchmark is both comprehensive and practically meaningful.

\section{Experiments}
\subsection{Experimental Setup}
\noindent \textbf{Implementation Details.} 
We sample 10,000 documents from the Huatuo26M\_encyclopedia dataset as the unlabeled corpus. In our training framework, we utilize Qwen2-7B-Instruct~\cite{yang2024qwen2} as the generator and BGE-Large-zh-v1.5~\cite{xiao2024c} as the retriever. Unless otherwise stated, all experiments are conducted under this Qwen+BGE configuration. Model training and evaluation are conducted on up to 5 NVIDIA A100 GPUs, each equipped with 40GB of memory. For fine-tuning the LLM, we employ the AdamW optimizer~\cite{loshchilov2017decoupled} in conjunction with a cosine learning rate scheduler. Training is executed for 1 epoch with a learning rate of 1e-5 and a batch size of 2. We set 200 warmup steps and configure the LoRA rank to 8. Retriever fine-tuning also uses the AdamW optimizer, with a linear decay schedule and an initial learning rate of 1e-5. The batch size per GPU is set at 4, and the maximum input sequence length is limited to 512. We apply a temperature of 0.02 and mine 7 hard negatives for each query to enhance training difficulty.

\noindent \textbf{Evaluation Settings.} For simplicity, we employ the LLM to generate a single hypothetical document for each query. The retrieval model embeds all queries, hypothetical documents, and corpus documents, with similarity scores calculated using cosine similarity. Documents in the corpus are ranked for each query based on these scores, and nDCG@10 is used as the primary evaluation metric to assess retrieval effectiveness. We set the temperature of LLM to 0.7 and repeat five times with different random seeds.

\noindent \textbf{Baseline Models.} 
To comprehensively evaluate CMIRB, we select several popular retrieval models. These include lexical retriever BM25~\cite{robertson2009probabilistic}; dense retrieval models such as Text2Vec-Large-Chinese~\cite{Text2vec}, PEG~\cite{wu2023towards}, BGE-Large-zh-v1.5~\cite{xiao2024c}, GTE-Large-zh~\cite{li2023towards}, and Piccolo-Large-zh~\cite{piccolo}; multilingual retrievers like mContriever (masmarco)~\cite{izacard2021unsupervised}, M3E-Large~\cite{Moka}, mE5 (multilingual-e5-large)~\cite{wang2024multilingual}; and text-embedding-ada-002~\cite{openaiemb}.

\subsection{Main Results}

The experimental results for various retrieval models, including SL-HyDE, on the CMIRB benchmark are presented in Table~\ref{tab:WHOLE}. We make the following observations.

(1) BM25 remains highly competitive in specific medical tasks. As a lexical retriever, it ranks documents based on TF-IDF matching scores calculated between queries and documents. Despite underperforming on the overall CMIRB benchmark, it displays strong results in tasks like medical news retrieval (78.9 vs. 73.33 for BGE) and medical post retrieval (66.95 vs. 67.13 for BGE). This can be attributed to the higher keyword overlap in datasets.

(2) No single retrieval model achieves optimal performance across all ten tasks. PEG and GTE each deliver the best performance on four datasets, while BGE and mE5 each excel in achieving the top results on one dataset. Dense models with better performance often utilize contrastive learning, pretraining on large-scale unlabeled data followed by fine-tuning on labeled data. Variations in training data distribution influence model effectiveness across different datasets, suggesting the need for specialized approaches.

(3) SL-HyDE consistently outperformed HyDE across all ten datasets. While HyDE shows slight overall improvements over BGE, it excels in medical knowledge retrieval but underperforms in medical consultation tasks. This discrepancy could be due to LLM's stronger handling of encyclopedia-type knowledge compared to the nuanced domain of patient-doctor consultations. In contrast, SL-HyDE achieved improvements over HyDE in all tasks, owing to its self-learning mechanism, which effectively enhances medical knowledge integration within both the generator and the retriever, while also aligning the outputs of the two models.

\subsection{Performance Analysis}
\noindent \textbf{Effect of Different Generators.} 
In Table~\ref{tab:llms}, we present SL-HyDE's performance with alternative fine-tuned LLMs as the generator, such as ChatGLM3-6B~\cite{team2024chatglm} and Llama2-7b-Chat~\cite{touvron2023llama}. 

Both models demonstrate performance improvements under SL-HyDE compared to HyDE. For instance, we observe a 4.65\% improvement with ChatGLM3 and an 8.23\% improvement with the Llama2 model. However, for Llama2, HyDE shows a slight decline compared to BGE. This is likely due to the fact that the pseudo-documents generated by the English-based Llama2 contained English content, which the downstream BGE retriever struggled to encode effectively. After fine-tuning, SL-HyDE improves by approximately 8\%, attributed to both the reduction of English content and the enhanced retriever's ability to encode medical knowledge, illustrating SL-HyDE's adaptability.

\begin{table}
  \resizebox{\linewidth}{!}{
  \begin{tabular}{l|c|c|c|c|c|l}
  \toprule
     Task   &Know. &Consu. &News & Post & Literature &Avg.(All) \\
     \hline
     \multicolumn{7}{c}{\textbf{ChatGLM3 as Generator + BGE as Retriever}} \\
     \hline
    HyDE &62.43 & 46.43 & 73.89 & 70.88 & 44.46 & 56.02 \\
    SL-HyDE &66.26 & 48.55 & 76.78 & 72.29 & 46.40 & 58.63 \\
    Improve. &$\uparrow$ \textbf{6.14\%} & $\uparrow$ \textbf{4.57\%} & $\uparrow$ \textbf{3.91\%} & $\uparrow$ \textbf{1.99\%} & $\uparrow$ \textbf{4.36\%} & $\uparrow$ \textbf{4.65\%} \\
    \hline
    \multicolumn{7}{c}{\textbf{Llama2 as Generator + BGE as Retriever}} \\
     \hline
    HyDE &55.74 & 40.62 & 72.90 & 72.22 & 45.30 & 52.48 \\
    SL-HyDE &63.66 & 45.44 & 77.17 & 71.99 & 45.75 & 56.80 \\
    Improve.  &$\uparrow$ \textbf{14.21\%} & $\uparrow$ \textbf{11.87\%} & $\uparrow$ \textbf{5.86\%} & $\downarrow$ \textbf{0.32\%} & $\uparrow$ \textbf{0.99\%} & $\uparrow$ \textbf{8.23\%} \\

\bottomrule
\end{tabular}
}
\caption{Performance of different generators.}
\label{tab:llms}
\end{table}

\noindent \textbf{Effect of Different Retrievers.}
We consider fine-tuning the other two retrieval models: PEG which achieves optimal performance on CMIRB, and a multilingual retriever mE5. 

In Table~\ref{tab:retrievals}, we observe that the standard HyDE method offers some improvement over using only the retriever, but the overall performance is significantly enhanced with the application of SL-HyDE. For example, the top-performing PEG model on the CMIRB benchmark improved from 57.46\% to 60.97\%, representing a substantial increase in retrieval tasks. This underscores SL-HyDE's ability to boost retrieval performance across various retriever models.

\begin{table}
  \resizebox{\linewidth}{!}{
  \begin{tabular}{l|c|c|c|c|c|l}
  \toprule
     Task   &Know. &Consu. &News & Post & Literature &Avg.(All) \\
     \hline
     \multicolumn{7}{c}{\textbf{Qwen2 as Generator + mE5 as Retriever}} \\
     \hline
    HyDE &65.77 & 43.15 & 75.92 & 68.15 & 38.58 & 54.80 \\
    SL-HyDE &68.60 & 44.83 & 77.59 & 66.81 & 42.33 & 56.94 \\
    Improve. &$\uparrow$ \textbf{4.31\%} & $\uparrow$ \textbf{3.90\%} & $\uparrow$ \textbf{2.20\%} & $\downarrow$ \textbf{1.97\%} & $\uparrow$ \textbf{9.72\%} & $\uparrow$ \textbf{3.90\%} \\
    \hline
    \multicolumn{7}{c}{\textbf{Qwen2 as Generator + PEG as Retriever}} \\
     \hline
    HyDE &66.03 & 49.73 & 80.49 & 72.51 & 38.87 & 57.80 \\
    SL-HyDE &69.96 & 50.97 & 80.89 & 75.93 & 45.03 & 60.97 \\
    Improve. &$\uparrow$ \textbf{5.96\%} & $\uparrow$ \textbf{2.50\%} & $\uparrow$ \textbf{0.50\%} & $\uparrow$ \textbf{4.72\%} & $\uparrow$ \textbf{15.86\%} & $\uparrow$ \textbf{5.48\%} \\

\bottomrule
\end{tabular}
}
\caption{Performance of different retrievers.}
\label{tab:retrievals}
\end{table}

\begin{table}
  \resizebox{\linewidth}{!}{
  \begin{tabular}{l|c|c|c|c|c|l}
  \toprule
     Task   &Know. &Consu. &News & Post & Literature &Avg.(All) \\
     \hline
    SL-HyDE &69.26 & 49.26 & 76.95 & 73.81 & 43.75 & 59.38 \\
   \ \ w/ D.  &68.00 & 41.86 & 71.94 & 68.02 & 37.36 & 54.43 \\
   \ \ w/ con. &69.04 & 45.51 & 73.38 & 69.53 & 44.81 & 57.62 \\
    \ \ w/ K-D. &69.30 & 50.17 & 77.38 & 74.55 & 45.42 & 60.12 \\
\bottomrule
\end{tabular}
}
\caption{Performance of different fusing strategies.}
\label{tab:pseudo}
\end{table}

\noindent \textbf{Effect of Different Fusing Strategies.} 
In this section, we test several methods for incorporating hypothetical documents. SL-HyDE: This method encodes the original query and the hypothetical documents separately, then applies mean pooling to obtain the final query vector. SL-HyDE w/ D: Only the hypothetical document is used as the query for retrieval. SL-HyDE w/ con: The original query and the hypothetical document are concatenated into a single string to form a new query. SL-HyDE w/ K-D: This approach generates five documents. 

Table~\ref{tab:pseudo} shows that the combination of the original query and hypothetical documents is optimal. Sole reliance on hypothetical documents significantly reduces performance, especially in medical consultation tasks, where original queries contain critical information. The string concatenation method introduces some performance degradation, indicating that the generated documents may contain noise at the string level, whereas average pooling effectively mitigates it. Generating multiple hypothetical documents increases coverage and improves performance across tasks. However, it often leads to a K-fold increase in inference time. Therefore, we need to balance efficiency and accuracy to select the number of hypothetical documents.

\subsection{Ablation Study} To further analyze the gains brought by the internal architecture of SL-HyDE, we conduct two sets of ablation experiments: (1) SL-HyDE w/o BGE-FT, which uses the fine-tuned LLM as the generator and the raw BGE as the retriever; (2) SL-HyDE w/o Qwen-FT, which utilizes raw LLM as the generator and the fine-tuned BGE as the retriever. 

Table~\ref{tab:Ablation} demonstrates that fine-tuning both components substantially enhances performance, validating the efficacy of the self-learning mechanism. Notably, fine-tuning the retriever yields greater gains, suggesting that BGE benefits significantly from domain-specific adaptation. However, our approach fine-tunes both the retriever and the generator, boosting their performance between the two to enhance retrieval tasks.

\begin{table}
  \resizebox{\linewidth}{!}{
  \begin{tabular}{l|c|c|c|c|c|l}
  \toprule
     Task   &Know. &Consu. &News & Post & Literature &Avg.(All) \\
     \hline
    HyDE &63.70 & 47.63 & 74.32 & 73.07 & 42.42 & 56.62 \\
    SL-HyDE &\textbf{69.26} & \textbf{49.26} &\textbf{76.95} & 73.81 & \textbf{43.75} &\textbf{59.38} \\
    \ \  w/o BGE-FT &64.32 & 47.95 & 74.87 & 72.91 & 43.24 & 57.11 \\
    \ \ w/o Qwen-FT &68.75 & 48.85 & 76.63 &\textbf{74.52} & 43.11 & 58.77 \\

\bottomrule
\end{tabular}
}
\caption{Performance of different variants.}
\label{tab:Ablation}
\end{table}

\subsection{Case Study}
To intuitively show how the SL-HyDE makes a difference in the hypothetical documents and retrieval performance, we present examples in Table~\ref{tab:Case} to compare the hypothetical document generated by HyDE and SL-HyDE. The query is \textit{How to treat a hernia?}. While HyDE generates a general document discussing \textit{conservative and surgical treatments}, it lacks specificity for different patient groups. In contrast, SL-HyDE produces a document mentioning \textit{hernia belts for infants and elderly patients}, closely matching the target document’s details. This improved relevance led to a higher retrieval ranking ($2nd$ vs. $10th$), demonstrating how more precise hypothetical documents enhance retrieval performance.

\begin{table}
  \resizebox{\linewidth}{!}{
  \begin{tabular}{l}
  \toprule
  \textbf{Query: } How to treat a hernia?\\
  \textbf{Target Doc:} Inguinal Hernia Treatment Plan. For conventional treatment, \\ a 1-year-old infant can use a hernia belt for compression. As the muscles \\ gradually strengthen, there may be a possibility of spontaneous recovery. \\ For elderly and frail individuals a hernia belt can be worn, but for \\ other patients, surgery is generally recommended...\\
  \hline
  \textbf{HyDE: } Hernia is a common disease caused by a weak area in the \\  abdominal wall, \textcolor{blue}{Treatment usually includes conservative and surgical} \\ \textcolor{blue}{methods}. For most patients, especially young and healthy individuals, \\ surgery is the preferred option... \textcolor{red}{(Rank: 10)}\\
  \textbf{SL-HyDE: } Hernia is a common condition that typically occurs... \textcolor{blue}{For} \\ \textcolor{blue}{infants},... \textcolor{blue}{the use of a hernia belt} to apply localized pressure can help\\ alleviate symptoms and promote the development of the abdominal \\ muscles,... \textcolor{blue}{For elderly or frail patients}, or those with severe underlying \\conditions,... \textcolor{blue}{wearing a hernia belt can help manage symptoms} and\\ reduce the risk of the hernia progressing further... \textcolor{red}{(Rank: 2)}\\

\bottomrule
\end{tabular}
}
\caption{The case study comparing with baseline.}
\label{tab:Case}
\end{table}

\section{Conclusions}
In this paper, we introduce an automated framework for zero-shot medical information retrieval, named SL-HyDE, which operates without the need for relevance labels. Utilizing an unlabeled medical corpus, we employ a self-learning, end-to-end training framework where the retriever guides the generator's training, and the generator, in turn, enhances the retriever. This process integrates medical knowledge to create hypothetical documents that are more effective in retrieving target documents. Furthermore, we present a comprehensive Chinese medical information retrieval benchmark, evaluating mainstream retrieval models against this new standard. Experimental findings demonstrate that SL-HyDE consistently improves retrieval accuracy over HyDE across ten datasets. Additionally, SL-HyDE shows strong adaptability and scalability, effectively enhancing retrieval performance across various combinations of generators and retrievers. In future work, we will extend SL-HyDE to other data-scarce domains to further evaluate its generalizability across different settings. In addition, we will explore reinforcement learning to train more capable retrievers and enhance reasoning in complex medical retrieval tasks. 

\section{Limitations}
While our work effectively addresses the adaptation challenges of HyDE in low-resource scenarios, several limitations remain. First, our study primarily focuses on the medical domain and provides a preliminary exploration in the legal domain (see Appendix~\ref{app:more_experiment}), but we have not extended our investigation to other vertical domains such as economics or education. Second, although we experiment with three open-source LLMs, Qwen2, LlaMA2, and ChatGLM3, as generators, we do not include more recent or diverse model families such as Qwen3 or Gemini, which may exhibit different generation behaviors. Third, our data construction pipeline relies on LLMs for query-document matching and pseudo-relevant pair filtering. The effectiveness of these components depends on the model's instruction-following ability and its sensitivity to domain-specific nuances, which may introduce hallucinations or spurious correlations.

% Bibliography entries for the entire Anthology, followed by custom entries
%\bibliography{anthology,custom}
% Custom bibliography entries only
\bibliography{main}

\clearpage
\appendix
\begin{figure*}[t!]
% \begin{table*}[t!]
\setlength{\tabcolsep}{3pt}
\centering
\resizebox{\textwidth}{!}{
\begin{tabular}{l|c|c}
\toprule 
Model &Size & Model Link \\
\midrule
\multicolumn{3}{c}{\textbf{Retrieval Models}} \\
\midrule
BM25~\cite{robertson2009probabilistic} &N/A &~\url{https://github.com/castorini/pyserini} \\
Text2Vec~\cite{Text2vec}&325M & \url{https://huggingface.co/GanymedeNil/text2vec-large-chinese} \\
PEG~\cite{wu2023towards} &335M &\url{https://huggingface.co/TownsWu/PEG}  \\
BGE~\cite{xiao2024c} &335M & \url{https://huggingface.co/BAAI/bge-large-zh-v1.5} \\
GTE~\cite{li2023towards} &335M & \url{https://huggingface.co/thenlper/gte-large-zh}\\
Piccolo~\cite{piccolo} &335M & \url{https://huggingface.co/sensenova/piccolo-large-zh} \\
Contriever~\cite{izacard2021unsupervised} &109M &\url{https://huggingface.co/facebook/mcontriever-msmarco} \\
M3E~\cite{Moka}&340M & \url{https://huggingface.co/moka-ai/m3e-large} \\
mE5~\cite{wang2024multilingual}  &560M& \url{https://huggingface.co/intfloat/multilingual-e5-large} \\
OpenAI-Ada-002~\cite{openaiemb}&N/A & \url{https://openai.com/index/new-and-improved-embedding-model/} \\
\midrule
\multicolumn{3}{c}{\textbf{Large Language Models}} \\
\midrule
Qwen2~\cite{yang2024qwen2}  &7B &\url{https://huggingface.co/Qwen/Qwen2-7B-Instruct} \\
Llama2~\cite{touvron2023llama} &7B  & \url{https://huggingface.co/meta-llama/Llama-2-7b-chat-hf} \\
ChatGLM3~\cite{team2024chatglm} &7B &\url{https://huggingface.co/THUDM/chatglm3-6b} \\
\bottomrule
\end{tabular}
}
\captionof{table}{Detailed information on all of the retrieval models and large language models in our paper.}
\label{tab:model_deatails}
\end{figure*}

\section{Models}
\subsection{Baselines}
To comprehensively evaluate the performance of existing retrievers on CMIRB, we selected 10 representative models, all of which have achieved strong results on the MTEB leaderboard\footnote{\href{https://huggingface.co/spaces/mteb/leaderboard}{https://huggingface.co/spaces/mteb/leaderboard}}. For details regarding the retrievers and large reasoning models evaluated throughout the paper, please refer to Table~\ref{tab:model_deatails}.

\noindent \textbf{BM25}~\cite{robertson2009probabilistic}. \ \ BM25 is a commonly used baseline retriever which uses bag-of-words and TF-IDF to perform lexical retrieval. In this paper, BM25 is implemented with Pyserini~\cite{lin2021pyserini} using the default hyperparameters to index snippets from all corpora.

\noindent \textbf{Text2Vec}~\cite{Text2vec}. \ \ It is a cosine sentence model based on a linguistically-motivated pre-trained language model (LERT).

\noindent \textbf{PEG}~\cite{wu2023towards}. \ \  Wu et al.,~\cite{wu2023towards} proposes the PEG, which is trained on more than 100 million data, encompassing a wide range of domains and covering various tasks.

\noindent \textbf{BGE}~\cite{xiao2024c}. \ \  It takes a compound recipe to train general-purpose text embedding, including, embedding-oriented pre-training, contrastive learning with sophisticated negative sampling, and instruction-based fine-tuning.

\noindent \textbf{GTE}~\cite{li2023towards}. \ \  It presents a multi-stage contrastive learning approach to develop text embedding model that can be applied to various tasks.

\noindent \textbf{Piccolo}~\cite{piccolo}. \ \  Piccolo is a general-purpose Chinese embedding model trained using a two-stage process with weakly supervised and manually labeled text pairs.

\noindent \textbf{Contriever }~\cite{izacard2021unsupervised}. \ \  It is a multilingual dense retriever with contrastive learning, which fine-tunes the pre-trained mContriever model on MS MARCO dataset.

\noindent \textbf{M3E}~\cite{Moka}. \ \  M3E (Moka Massive Mixed Embedding) is a bilingual text embedding model trained on over 22 million Chinese sentence pairs, supporting tasks like cross-lingual text similarity and retrieval.

\begin{table}
  \resizebox{\linewidth}{!}{
  \begin{tabular}{l}
  \toprule
\textbf{Q2P Prompt} \\
   \hline
 Please generate a medical content paragraph to answer\\ this question. \\
 Question: {QUESTION} \\
 Paragraph:  \\
 \hline
   \textbf{T2P Prompt} \\
   \hline
 Please generate a medical content paragraph based on \\this title. \\
 Title: {TITLE} \\
 Paragraph:  \\
  \hline
    \textbf{P2P Prompt} \\
    \hline
Please generate a similar medical paragraph for the \\following text. \\
 Text: {TEXT} \\
 Similar Paragraph:  \\
 
\bottomrule
\end{tabular}
}
\caption{Evaluation prompts for generators.}
  \label{tab:eva-prompts}
\end{table}

\noindent \textbf{mE5}~\cite{wang2024multilingual}. \ \ Multilingual E5 text embedding models that are trained with a multi-stage pipeline, involving contrastive pre-training on 1 billion multilingual text pairs, and fine-tuning on labeled datasets.

\noindent \textbf{OpenAI-Ada-002}~\cite{openaiemb}. \ \ It is a highly efficient text embedding model that converts natural language into dense vectors for a wide range of applications, including semantic search and similarity tasks.

For the generator, we selected three highly powerful large language models.

\noindent \textbf{Qwen2}~\cite{yang2024qwen2}. \ \ Qwen2 is a comprehensive suite of foundational and instruction-tuned language models, encompassing a parameter range from 0.5 to 72 billion, featuring dense models and a Mixture-of-Experts model.

\noindent \textbf{ChatGLM3}~\cite{team2024chatglm}. \ \ ChatGLM3-6B is a next-generation conversational pre-trained model with strong performance across tasks like semantics, reasoning, and code execution, and supports complex scenarios such as tool use and function calls.

\noindent \textbf{Llama2}~\cite{touvron2023llama}. \ \  Llama2 is an auto-regressive language model that uses an optimized transformer architecture. The tuned versions utilize supervised fine-tuning (SFT) and reinforcement learning with human feedback (RLHF) to align with human preferences for helpfulness and safety.

% \subsection{Training Details}
% \label{sec: appendix-implementation}
% We run model training and model evaluation on up to 5 NVIDIA A100 GPUs with 40GB memory. For fine-tuning LLMs and retrievers, we list the hyperparameters in Table~\ref{tab:Implementation}.

\subsection{Evaluation Settings}
\label{app:evaluation}
We use the C-MTEB\footnote{\href{https://github.com/FlagOpen/FlagEmbedding/tree/master/research/C_MTEB}{C-MTEB}} framework to evaluate the performance of various retrieval models on CMIRB. To ensure stability, we set the temperature of LLM to 0.7 and repeat five times with different random seeds. For each dataset, the prompts used to generate pseudo-documents are shown in Figure~\ref{tab:eva-prompts}. The IIYIPost and CSLCite datasets utilize the T2P template to prompt LLMs to generate documents based on the given title. For the CSLRel dataset, we employ the P2P template to instruct the model to produce similar text. As for the other datasets, the Q2P template is employed by the LLM to generate answers to medical questions.

\subsection{SL-HyDE vs. HyDE}
Our approach, SL-HyDE, builds upon HyDE~\cite{gao2022precise} with several enhancements while retaining some similarities. First, both SL-HyDE and HyDE follow the same inference process. Each uses a large model to generate a hypothetical document based on the query, which the retriever then employs to locate the most relevant document. Second, neither SL-HyDE nor HyDE requires labeled data, which allows for rapid deployment. HyDE is especially advantageous in real-world scenarios where efficient retrieval can be executed simply by selecting a generator and a retriever. However, for tasks needing domain-specific knowledge, such as medical information retrieval, deploying HyDE directly may not yield optimal results. One potential strategy is to fine-tune the generator and retriever separately using labeled medical data before deploying the HyDE framework. The primary challenge here in acquiring labeled data, and fine-tuning the models separately often leads to suboptimal performance.

SL-HyDE improves upon this by integrating a self-learning mechanism, transforming HyDE into a trainable end-to-end framework. This mechanism enables both the generator and the retriever to better adapt to the medical domain. Supervision signals for the generator's training are derived from the retriever, and vice versa, facilitating mutual enhancement through this self-learning process. This holistic approach results in improved performance in retrieval tasks. Overall, SL-HyDE offers an efficient and convenient solution for enhancing HyDE's performance in the medical domain, particularly when dealing with unlabeled corpora.

\subsection{More Experiment Results}
\label{app:more_experiment}
Table~\ref{tab:Recall_100} presents the performance of 10 retrieval models on CMIRB in terms of Recall@100. In Table~\ref{tab:combination}, we present a more detailed breakdown of the performance of various LLM and retriever combinations across the 10 datasets. 

SL-HyDE can be easily applied to other domains that lack labeled data. By fine-tuning both the generator and retriever using only a small amount of unstructured domain text, it builds an effective retrieval system. Specifically,  we apply SL-HyDE to the English legal domain. We sample 10k law texts from pile-of-law \footnote{\href{https://huggingface.co/datasets/pile-of-law/pile-of-law}{https://huggingface.co/datasets/pile-of-law/pile-of-law}} and use Llama-2-7b-chat-hf as the generator and BGE-Large-en-V1.5 as the retriever. We evaluate three information retrieval datasets in the law domain from MTEB. The results in Table~\ref{tab:legal} shows that SL-HyDE (77.25\%) significantly outperforms HyDE (75.52\%) in the legal domain.

\begin{table*}[t!]
  \resizebox{\linewidth}{!}{
  \begin{tabular}{l|ccc|ccc|c|c|cc|l}
  \toprule
     Task   &\multicolumn{3}{c|}{Knowledge Retrieval} &\multicolumn{3}{|c|}{Consulation Retrieval} &News & Post & \multicolumn{2}{|c|}{Literature Retrieval} & \\
     \hline
     Dataset &MedExam &DuBaike &DXYDis. &Medical &Cmedqa &DXYCon. &Covid &IIYiPost & CSLCite &CSLRel &Average \\
     \hline
    BM25 & 75.61 & 56.92 & 72.91 & 44.20 & 17.26 & 37.33 & 96.47 & 89.98 & 67.19 & 72.66 & 63.05 \\
    Text2Vec(large) & 89.81 & 79.25 & 78.01 & 52.80 & 42.99 & 64.58 & 88.83 & 74.78 & 61.96 & 70.39 & 70.34 \\
    mContriever & 93.40 & 86.48 & 84.06 & 61.50 & 53.40 & 62.67 & 84.93 & 70.72 & 72.25 & 84.97 & 75.44 \\
    mE5(large) & 93.83 & 98.43 & 96.02 & 70.90 & 57.95 & 80.38 & 97.05 & 91.64 & 77.31 & 91.12 & 85.46 \\
    M3E(large) & 86.08 & 98.43 & 93.55 & 74.00 & 70.61 & 86.96 & 93.26 & 88.97 & 76.09 & \textbf{96.58} & 86.45 \\
    GTE(large) & 87.52 & 96.54 & 95.86 & \textbf{87.00} & \textbf{84.95} & 89.50 & \textbf{99.47} & 93.41 & \textbf{83.25} & \textbf{96.58} & 91.41 \\
    piccolo(large) & 89.67 & 99.06 & 96.81 & 82.80 & 84.81 & 91.09 & \textbf{99.47} & 95.69 & 83.07 & 92.25 & 91.47 \\
    PEG(large) & 95.41 & \textbf{98.74} & \textbf{98.01} & 83.70 & 84.64 & 89.50 & 98.74 & \textbf{96.83} & 81.15 & 92.25 & \textbf{91.90} \\
    BGE(large) & \textbf{97.42} & \textbf{98.74} & 96.81 & 81.20 & 82.57 & \textbf{91.30} & 98.10 & 95.69 & 80.80 & 96.36 & \textbf{91.90} \\
\bottomrule
\end{tabular}
}
\caption{Performance of various Retrieval models on CMIRB benchmark. All scores denote Recall@100. The best score on a given dataset is marked in bold.}
  \label{tab:Recall_100}
\end{table*}

\begin{table*}[t!]
  \resizebox{\linewidth}{!}{
  \begin{tabular}{l|ccc|ccc|c|c|cc|l}
  \toprule
     Task   &\multicolumn{3}{c|}{Knowledge Retrieval} &\multicolumn{3}{|c|}{Consulation Retrieval} &News & Post & \multicolumn{2}{|c|}{Literature Retrieval} & \\
     \hline
     Dataset &MedExam &DuBaike &DXYDis. &Medical &Cmedqa &DXYCon. &Covid &IIYiPost & CSLCite &CSLRel &Average \\
     \hline
   \multicolumn{12}{c}{\textbf{ChatGLM3 as Generator + BGE as Retriever}} \\
   \hline
    HyDE&61.96 & 54.25 & 71.07 & 56.32 & 37.73 & 45.23 & 73.89 & 70.88 & 45.11 & 43.80  & 56.02 \\
    SL-HyDE &67.12 & 59.40  & 72.25 & 57.16 & 38.77 & 49.71 & 76.78 & 72.29 & 45.81 & 46.98 &  58.63 \\
    Improve. &$\uparrow$ \textbf{8.33\%} & $\uparrow$ \textbf{9.49\%} & $\uparrow$ \textbf{1.66\%} & $\uparrow$ \textbf{1.49\%} & $\uparrow$ \textbf{2.76\%} & $\uparrow$ \textbf{9.90\%} & $\uparrow$ \textbf{3.91\%} & $\uparrow$ \textbf{1.99\%} & $\uparrow$ \textbf{1.55\%} & $\uparrow$ \textbf{7.26\%} & $\uparrow$ \textbf{4.65\%} \\
    \hline
    \multicolumn{12}{c}{\textbf{Llama2 as Generator + BGE as Retriever}} \\
   \hline
    HyDE &53.10  & 45.78 & 68.34 & 53.51 & 31.29 & 37.07 & 72.90  & 72.22 & 44.19 & 46.41 & 52.48 \\
     SL-HyDE &64.88 & 56.30  & 69.81 & 54.68 & 36.93 & 44.72 & 77.17 & 71.99 & 44.62 & 46.88 &  56.80 \\
     Improve. &$\uparrow$ \textbf{22.18\%} & $\uparrow$ \textbf{22.98\%} & $\uparrow$ \textbf{2.15\%} & $\uparrow$ \textbf{2.19\%} & $\uparrow$ \textbf{18.02\%} & $\uparrow$ \textbf{20.64\%} & $\uparrow$ \textbf{5.86\%} & $\downarrow$ \textbf{0.32\%} & $\uparrow$ \textbf{0.97\%} & $\uparrow$ \textbf{1.01\%} & $\uparrow$ \textbf{8.23\%} \\
    \hline
    \multicolumn{12}{c}{\textbf{Qwen2 as Generator + mE5 as Retriever}} \\
    \hline
    HyDE &65.18 & 56.35 & 75.77 & 54.31 & 32.02 & 43.12 & 75.92 & 68.15 & 45.66 & 31.50  & 54.80 \\
    SL-HyDE & 71.36 & 59.50  & 74.95 & 54.68 & 33.95 & 45.87 & 77.59 & 66.81 & 45.65 & 39.01 &  56.94 \\
    Improve. &$\uparrow$ \textbf{9.48\%} & $\uparrow$ \textbf{5.59\%} & $\downarrow$ \textbf{1.08\%} & $\uparrow$ \textbf{0.68\%} & $\uparrow$ \textbf{6.03\%} & $\uparrow$ \textbf{6.38\%} & $\uparrow$ \textbf{2.20\%} & $\downarrow$ \textbf{1.97\%} & $\downarrow$ \textbf{0.02\%} & $\uparrow$ \textbf{23.84\%} & $\uparrow$ \textbf{3.90\%} \\
    \hline
    \multicolumn{12}{c}{\textbf{Qwen2 as Generator + PEG as Retriever}} \\
    \hline
    HyDE & 64.87 & 55.04 & 78.18 & 58.47 & 41.47 & 49.25 & 80.49 & 72.51 & 43.56 & 34.17 & 57.80 \\
   SL-HyDE &72.04 & 60.26 & 77.59 & 59.81 & 40.43 & 52.68 & 80.89 & 75.93 & 47.53 & 42.53 &  60.97 \\
    Improve. &$\uparrow$ \textbf{11.05\%} & $\uparrow$ \textbf{9.48\%} & $\downarrow$ \textbf{0.75\%} & $\uparrow$ \textbf{2.29\%} & $\downarrow$ \textbf{2.51\%} & $\uparrow$ \textbf{6.96\%} & $\uparrow$ \textbf{0.50\%} & $\uparrow$ \textbf{4.72\%} & $\uparrow$ \textbf{9.11\%} & $\uparrow$ \textbf{24.47\%} & $\uparrow$ \textbf{5.48\%} \\
\bottomrule
\end{tabular}
}
\caption{Performance of different combinations of generators and retrievers on CMIRB benchmark.}
  \label{tab:combination}
\end{table*}

\begin{table}
  \resizebox{\linewidth}{!}{
  \begin{tabular}{l|l|l|l|l}
  \toprule
     Dataset   &legal\_ &legalbench\_ &legalbench\_ & Average \\
     &summar. &contracts\_qa &lobbying & \\
     \hline
    BGE &59.99	&73.52	&91.51	&75.01  \\
  HyDE  &58.95	&74.82	&92.78	&75.52 \\
   SL-HyDE &\textbf{63.50}	&\textbf{75.10}	&\textbf{93.15}	&\textbf{77.25} \\

\bottomrule
\end{tabular}
}
\caption{Performance of SL-HyDE in legal domain.}
\label{tab:legal}
\end{table}

\section{CMIRB Datasets}

\begin{table*}
  \resizebox{\linewidth}{!}{
  \begin{tabular}{lcccc}
  \toprule
    Dataset &Query URL  &\#Samples &Document URL  &\#Samples  \\
   \hline
 MedExam &\href{https://github.com/jind11/MedQA}{https://github.com/jind11/MedQA} & 3,426 &\href{https://github.com/jind11/MedQA}{https://github.com/jind11/MedQA} & 27,871\\
  DuBaike &\href{https://github.com/baidu/DuReader}{https://github.com/baidu/DuReader} & 20,000 &\href{https://baike.baidu.com/}{https://baike.baidu.com/} & 56,441\\
   DXYDisease &\href{https://dxy.com/diseases}{https://dxy.com/diseases} & 61,840 &\href{https://dxy.com/diseases}{https://dxy.com/diseases} & 61,840\\
    MedicalRetrieval &\href{https://huggingface.co/datasets/C-MTEB/MedicalRetrieval}{https://huggingface.co/datasets/C-MTEB/MedicalRetrieval} & 1,000 &\href{https://huggingface.co/datasets/C-MTEB/MedicalRetrieval}{https://huggingface.co/datasets/C-MTEB/MedicalRetrieval} & 100,999\\
 CmedqaRetrieval &\href{https://huggingface.co/datasets/C-MTEB/CmedqaRetrieval}{https://huggingface.co/datasets/C-MTEB/CmedqaRetrieval} & 3,999 &\href{https://huggingface.co/datasets/C-MTEB/CmedqaRetrieval}{https://huggingface.co/datasets/C-MTEB/CmedqaRetrieval} & 100,001\\
 DXYConsult &\href{https://dxy.com/questions/}{https://dxy.com/questions/} & 13,057 &\href{https://dxy.com/questions/}{https://dxy.com/questions/} & 13,057\\
  CovidRetrieval &\href{https://huggingface.co/datasets/C-MTEB/CovidRetrieval}{https://huggingface.co/datasets/C-MTEB/CovidRetrieval} &949 &\href{https://huggingface.co/datasets/C-MTEB/CovidRetrieval}{https://huggingface.co/datasets/C-MTEB/CovidRetrieval} & 100,001\\
   IIYiPost &\href{https://bbs.iiyi.com/}{https://bbs.iiyi.com/} & 37,065 &\href{https://bbs.iiyi.com/}{https://bbs.iiyi.com/} & 37,065\\
    CSLCite &\href{https://github.com/ydli-ai/CSL}{https://github.com/ydli-ai/CSL} & 934 &\href{https://med.wanfangdata.com.cn/}{https://med.wanfangdata.com.cn/} & 36,783\\
     CSLRel &\href{https://github.com/ydli-ai/CSL}{https://github.com/ydli-ai/CSL} & 934 &\href{https://med.wanfangdata.com.cn/}{https://med.wanfangdata.com.cn/} & 36,783\\

\bottomrule
\end{tabular}
}
\caption{Dataset collection sources and quantity statistics.}
  \label{tab:data-collection}
\end{table*}

\subsection{Data Process}
\label{sec: appendix-data}
\begin{figure}[t!]
\centering
  \includegraphics[width=\linewidth]{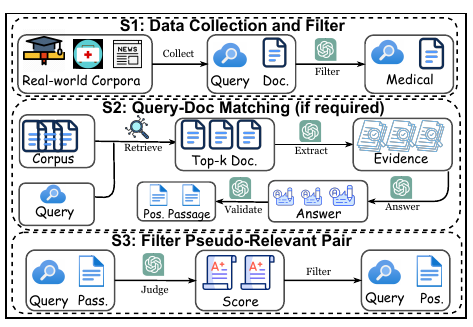}
  \caption{CMIRB benchmark construction pipeline.}
  \label{fig: data_pipeline}
\end{figure}

We curated a substantial dataset from various medical resources, as presented in Table~\ref{tab:data-collection}, which details the source distribution and data volume. Our data preprocessing pipeline, depicted in Figuer~\ref{fig: data_pipeline} and Algorithm~\ref{alg:data}, employs prompt templates outlined in Figure~\ref{fig: prompt_1} and Figure~\ref{fig:prompt_2}.

Initially, we use ChatGPT\footnote{\href{https://openai.com/chatgpt}{https://openai.com/chatgpt}} to perform medical relevance detection on the texts, eliminating non-medical content (lines 3-8). Subsequently, ChatGPT assesses query-document relevance, filtering out low-relevance examples (lines 27-33). Our relevance assessment considers semantic alignment and the practical significance of data samples for their respective tasks, as highlighted in prompt~\ref{tab:data-prompts2}.

For the MedExam and DuBaike datasets, the direct query-document signal isn't initially provided. Both queries and documents in the MedExam dataset originate from Work~\cite{jin2021disease}, where 100 randomly selected questions have corpus documents containing evidence sufficient to

{\scriptsize
\begin{algorithm}[H]
\caption{Data Preprocessing Pipeline}
\label{alg:data}
\begin{algorithmic}[1]
\State \textbf{Input:} Query set $Q$, Document set $D$, A large language model $\mathrm{LLM}$ (e.g., ChatGPT)
\State \textbf{Output:} High-quality, highly relevant query-document pair collection

\State // Step 1: Filter out medically irrelevant 
\For{each query $q \in Q$, $d \in D$}
    \State $med_{score} \gets \mathrm{LLM}.\text{med\_score}(q/d)$
    \If {$med_{score} < threshold$} 
        \State Remove $q/d$ 
    \EndIf
\EndFor

\State // Step 2: Matching positive pairs 
\If{query-document matching}
    \For{each query $q \in Q$}
        \State // Retrieve top-k documents
        \State $D_k \gets \text{BM25}(q, D)$
        \State $D_k \gets \mathrm{LLM}.\text{reranking}(q; D_k)$

        \State // Extract evidence snippets
        \State $E_k \gets \mathrm{LLM}.\text{extract\_evidence}(q, D_k)$
        
        \State // Generate answers
        \State $A_k \gets \mathrm{LLM}.\text{answer}(q, E_k)$
        
        \For{each document $d_i$}
            \If{$\mathrm{LLM}.\text{validate}(a_i, d_i)$}
                \State Store $(q, d_i)$ 
            \EndIf
        \EndFor
    \EndFor
\EndIf

\State // Step 3: Filter out pseudo-relevant pairs
\For{each matched pair $(q, d)$}
    \State $rel_{score} \gets \mathrm{LLM}.\text{filter\_score}(q, d)$
    \If {$rel_{score} < threshold$} 
        \State Remove $(q, d)$ 
    \EndIf
\EndFor
\end{algorithmic}
\end{algorithm}
}

\noindent answer them, verified manually by the authors. In the DuBaike dataset, queries from Baidu Search and Baidu Zhidao often match the content distribution of Baidu Baike. These factors allow us to design a query-matching algorithm to locate the valuable document.

We leverage ChatGPT's capabilities to identify the most relevant documents. Starting with a query, we use the BM25 to retrieve the top 20 relevant documents, which GPT then ranks to identify the top 3 most relevant. Ideally, these documents should be semantically related and provide sufficient answers or evidence for the query. Therefore, ChatGPT extracts document segments as evidence details for the query.

To verify the sufficiency of this evidence, GPT generates an answer to the query based on the extracted evidence fragment. A self-verification step follows: if the GPT-generated answer aligns with the document, the document is deemed a positive match for the query. For MedExam, where queries are multiple-choice questions, we verify model answers against correct ones. For DuBaike, queries are medical knowledge questions, and answers are encyclopedic. GPT scores the generated and reference answers for consistency in expressing the same medical knowledge. This detailed process is outlined in lines 10-26.

Through this iterative loop of self-ranking, evidence searching, answering, and verification, combined with ChatGPT's advanced knowledge capabilities, we ensure high-quality, highly relevant query-document pairs.

\subsection{Data Example}
The datasets we constructed encompass various real-world medical scenarios, with examples from 10 different datasets illustrated in Table~\ref{tab:data_example_1} and Table~\ref{tab:data_example_2}. Queries can take the form of a medical paper title, a patient’s symptom description, or an exam question. Corresponding documents include abstracts of medical papers, doctor-patient diagnostic conversations, and reference materials for exam questions.

% \begin{table}
%   \caption{Implementation Details.}
%   \label{tab:Implementation}
%   % \resizebox{\linewidth}{!}{
%   \begin{tabular}{lcc}
%   \toprule
%     Hyperparameter &LLM  &Retrieval  \\
%    \hline
%  Learning rate &1e-5 & 1e-5 \\
%   Optimizer &AdamW & AdamW \\
%   Batch size &2 &4 \\
%  Epoch &1 &1 \\
%  Warmup steps &200 &- \\
%  Lora rank &8 &- \\
%  Query \& Passage length & -& 512 \& 512 \\
%  Temperature & -&0.02 \\
%  Hard negatives &- &7 \\
 
% \bottomrule
% \end{tabular}
% % }
% \end{table}

\begin{figure*}[htbp]
\centering
\begin{tcolorbox}[
    colframe = gray,       % 边框颜色
    colback = gray!5!white,             % 背景颜色
    coltitle = white,                   % 标题字体颜色
    coltext = black,                    % 文字颜色
    fonttitle = \bfseries,              % 标题字体加粗
    title = Medical Relevance Prompt,  % 标题内容
    boxrule = 1pt,                      % 边框宽度
    arc = 2mm,                          % 边角圆润
    width = \linewidth,                 % 宽度
    left = 7pt,                         % 左边距
    right = 7pt,                        % 右边距
    top = 5pt,                          % 上边距
    bottom = 5pt                        % 下边距
]
\fontsize{8.5pt}{10pt}\selectfont
You will receive a question-answer pair from Baidu Search. Your task is to evaluate whether the Q\&A is related to the medical field and output the result in JSON format. \\
 The JSON object must include the following keys: \\
- "reason": a string explaining the reason for your judgment. \\
- "label": an int, 0/1. \\
 Please adhere to the following steps: \\
- If the content mentioned in the question and answer includes medical information and is related to the medical field, the label should be 1. \\
- If most of the content in the question and answer is unrelated to the medical field, the label should be 0. \\
You need to make a judgment and provide a reason. Please output the result as required, and do not output any other content. \\
Here is the text: \\
Question: [QUESTION]  \\
Answer: [ANSWER]  \\
\end{tcolorbox}

\begin{tcolorbox}[
    colframe = gray,       % 边框颜色
    colback = gray!5!white,             % 背景颜色
    coltitle = white,                   % 标题字体颜色
    coltext = black,                    % 文字颜色
    fonttitle = \bfseries,              % 标题字体加粗
    title = Passage Reranking Prompt,  % 标题内容
    boxrule = 1pt,                      % 边框宽度
    arc = 2mm,                          % 边角圆润
    width = \linewidth,                 % 宽度
    left = 7pt,                         % 左边距
    right = 7pt,                        % 右边距
    top = 5pt,                          % 上边距
    bottom = 5pt                        % 下边距
]
\fontsize{8.5pt}{10pt}\selectfont
You will be given a medical question, a reference (standard) answer, and a model-generated answer. Your task is to evaluate the content similarity between the reference answer and the model-generated answer to determine whether they are conveying the same meaning. Your output is a JSON object, which must contain the following keys: \\
- "similarity\_score": a number between 0 and 1 indicating the content similarity between the two answers. \\
- "explanation": a detailed explanation of the similarities or differences that justify your similarity score. \\
 Please adhere to the following steps:  \\
- 1. Carefully read the medical question.\\
- 2. Review the reference answer and the model-generated answer.\\
- 3. Compare the two answers, focusing on content similarity—whether they convey the same meaning, and lead to the same conclusion. \\
- 4. Provide a similarity score between 0 and 1, where 1 indicates that the answers are identical in meaning, and 0 indicates different. \\
- 5. Justify your score by explaining the similarities or differences between the two answers. \\
The "explanation" should be in Chinese. and your output must always be a JSON object, do not output anything else.\\
Now here are the question, standard answer, and generated answer.\\
Question: [QUESTION] \\
Reference Answer: [REFERENCE ANSWER] \\
Model-generated Answer: [MODEL-GENERATED ANSWER] 
\end{tcolorbox}

\begin{tcolorbox}[
    colframe = gray,       % 边框颜色
    colback = gray!5!white,             % 背景颜色
    coltitle = white,                   % 标题字体颜色
    coltext = black,                    % 文字颜色
    fonttitle = \bfseries,              % 标题字体加粗
    title = Evidence Extracting Prompt,  % 标题内容
    boxrule = 1pt,                      % 边框宽度
    arc = 2mm,                          % 边角圆润
    width = \linewidth,                 % 宽度
    left = 7pt,                         % 左边距
    right = 7pt,                        % 右边距
    top = 5pt,                          % 上边距
    bottom = 5pt                        % 下边距
]
\fontsize{8.5pt}{10pt}\selectfont
You will be given a medical question, its answer and a related document. Your task is to extract evidence spans from the document that directly or indirectly support the answer to the medical question. Your output is a JSON object, which must contain the following keys: \\
- "evidence\_spans": a list, a list of passages. Please adhere to the following steps:  \\
- 1. Carefully read the medical question and its answer.\\
- 2. Review the content of the provided document. \\
- 3. Identify and extract the passage from the document that directly supports the correct answer to the question. \\
- 4. If no passage in the document can directly support the correct answer or answer the question, return an empty list.
\\
The "explanation" should be in Chinese. and your output must always be a JSON object, do not output anything else.\\
Here is the medical question, its answer, and the related document \\
Question: [QUESTION] \\
Answer: [ANSWER] \\
Document: [DOCUMENT] \\
\end{tcolorbox}

\caption{Prompt for data processing (I).}
\label{fig: prompt_1}
\end{figure*}

\begin{figure*}[htbp]
\centering
\begin{tcolorbox}[
    colframe = gray,       % 边框颜色
    colback = gray!5!white,             % 背景颜色
    coltitle = white,                   % 标题字体颜色
    coltext = black,                    % 文字颜色
    fonttitle = \bfseries,              % 标题字体加粗
    title = Answer by Evidence Prompt,  % 标题内容
    boxrule = 1pt,                      % 边框宽度
    arc = 2mm,                          % 边角圆润
    width = \linewidth,                 % 宽度
    left = 7pt,                         % 左边距
    right = 7pt,                        % 右边距
    top = 5pt,                          % 上边距
    bottom = 5pt                        % 下边距
]
\fontsize{8.5pt}{10pt}\selectfont
You will be given a medical exam question and one or more evidence spans that were extracted from related documents. Your task is to provide a detailed and comprehensive answer to the question based solely on the provided evidence spans. Your output is a JSON object, which must contain the following keys: \\
- "answer": a string, the answer you derive from the reference documents. \\
- "reason": a detailed explanation of your reasoning process leading to the answer. \\
 Please adhere to the following steps:  \\
- 1. Review the exam question.\\
- 2. Review the provided evidence spans.\\
- 3. Based solely on the information contained in the evidence spans, provide a detailed and comprehensive answer to the question. \\
- 4. If the evidence spans do not provide sufficient information to answer the question, state "The evidence passage can not answer the question." in "answer" and explain why. If you don't know the answer, don't guess. \\
You must not use any common knowledge, personal knowledge, or external information beyond the provided evidence spans. The "answer" and "reason" should be in Chinese. and your output must always be a JSON object, do not output anything else.\\
Now here are the exam question and reference documents. \\
Question: [QUESTION] \\
Evidence Spans: [EVIDENCE SPANS]
\end{tcolorbox}

\begin{tcolorbox}[
    colframe = gray,       % 边框颜色
    colback = gray!5!white,             % 背景颜色
    coltitle = white,                   % 标题字体颜色
    coltext = black,                    % 文字颜色
    fonttitle = \bfseries,              % 标题字体加粗
    title = Validate Answer Prompt,  % 标题内容
    boxrule = 1pt,                      % 边框宽度
    arc = 2mm,                          % 边角圆润
    width = \linewidth,                 % 宽度
    left = 7pt,                         % 左边距
    right = 7pt,                        % 右边距
    top = 5pt,                          % 上边距
    bottom = 5pt                        % 下边距
]
\fontsize{8.5pt}{10pt}\selectfont
You will be given a medical question, a reference (standard) answer, and a model-generated answer. Your task is to evaluate the content similarity between the reference answer and the model-generated answer to determine whether they are conveying the same meaning. Your output is a JSON object, which must contain the following keys: \\
- "similarity\_score": a number between 0 and 1 indicating the content similarity between the two answers. \\
- "explanation": a detailed explanation of the similarities or differences that justify your similarity score. \\
 Please adhere to the following steps:  \\
- 1. Carefully read the medical question.\\
- 2. Review the reference answer and the model-generated answer.\\
- 3. Compare the two answers, focusing on content similarity—whether they convey the same meaning, and lead to the same conclusion. \\
- 4. Provide a similarity score between 0 and 1, where 1 indicates that the answers are identical in meaning, and 0 indicates different. \\
- 5. Justify your score by explaining the similarities or differences between the two answers. \\
The "explanation" should be in Chinese. and your output must always be a JSON object, do not output anything else.\\
Now here are the question, standard answer, and generated answer.\\
Question: [QUESTION] \\
Reference Answer: [REFERENCE ANSWER] \\
Model-generated Answer: [MODEL-GENERATED ANSWER] 
\end{tcolorbox}

\begin{tcolorbox}[
    colframe = gray,       % 边框颜色
    colback = gray!5!white,             % 背景颜色
    coltitle = white,                   % 标题字体颜色
    coltext = black,                    % 文字颜色
    fonttitle = \bfseries,              % 标题字体加粗
    title = Query-Document Relevance Prompt,  % 标题内容
    boxrule = 1pt,                      % 边框宽度
    arc = 2mm,                          % 边角圆润
    width = \linewidth,                 % 宽度
    left = 7pt,                         % 左边距
    right = 7pt,                        % 右边距
    top = 5pt,                          % 上边距
    bottom = 5pt                        % 下边距
]
\fontsize{8.5pt}{10pt}\selectfont
You will be given a medical search query and its associated passage. Your task is to evaluate the quality of query-passage pairs intended for use in a medical encyclopedia knowledge retrieval evaluation dataset. Your output is a JSON object, which must contain the following keys: \\
- "quality\_score": an integer, a score from 1 to 5. \\
- "explanation": a string, providing a brief rationale for the given score. \\
 Please adhere to the following steps:  \\
- 1. Carefully read the query to understand the user's information need.\\
- 2. Review the passage to assess its relevance and targeted content in relation to the query.\\
- 3. Assign a quality score from 1 to 5 and explain your reasoning. \\
The "explanation" should be in Chinese. and your output must always be a JSON object, do not output anything else.\\
Now here are the query and passage.
Query: [QUERY] \\
Passage: [PASSAGE] \\
\end{tcolorbox}

\caption{Prompt for data processing (II).}
\label{fig:prompt_2}
\end{figure*}

\begin{table*}[!t]
    \centering
    \fontsize{9pt}{11pt}\selectfont
    \begin{tabular}{p{0.98\linewidth}}
    \midrule
        \rowcolor{gray!20}\textbf{MedExam} \\
    \midrule
    \begin{CJK}{UTF8}{gbsn}
        \textbf{Query:} 问题: 胃癌最常发生的转移途径是（）。选项: A:直接蔓延, B:血性转移, C:种植转移, D:淋巴转移, E:沿肠管转移.
    \end{CJK}\\
    \textit{\textbf{(EN)}} \textit{Question: The most common metastasis route for gastric cancer is (). Options: A: Direct spread, B: Hematogenous metastasis, C: Seeding metastasis, D: Lymphatic metastasis, E: Along the intestinal tract.}\\
      \begin{CJK}{UTF8}{gbsn}
        \textbf{Document:} 外科学 3.胃癌的扩散与转移 (2)淋巴转移：是胃癌的主要转移途径，进展期胃癌的淋巴转移率高达70\%左右，侵及黏膜下层的早期胃癌淋巴转移率近20\%。通常将引流胃的淋巴结分为16组，有的组还可以进一步分为若千亚组...
    \end{CJK}\\
    \textit{\textbf{(EN)}} \textit{Surgery 3. Gastric cancer dissemination and metastasis (2) Lymphatic metastasis: It is the primary route of metastasis for gastric cancer, with a lymphatic metastasis rate of about 70\% in advanced gastric cancer and approximately 20\% in early gastric cancer invading the submucosa. Lymph nodes draining the stomach are usually classified into 16 groups, with some groups further divided into several subgroups...}\\  
\midrule
        \rowcolor{gray!20}\textbf{DuBaike} \\
    \midrule
    \begin{CJK}{UTF8}{gbsn}
        \textbf{Query:} 强迫症的表现是什么?
    \end{CJK}\\
    \textit{\textbf{(EN)}} \textit{What are the manifestations of obsessive-compulsive disorder (OCD)?}\\
      \begin{CJK}{UTF8}{gbsn}
        \textbf{Document:} 强迫症 临床表现 多发人群焦虑症与遗传因素、个性特点、不良事件、应激因素等均有关系，尤其与患者的个性特点紧密相关，比如：过分追求完美、犹豫不决、谨小慎微、固执等，具备这些不良个性特征容易患强迫症...
    \end{CJK}\\
    \textit{\textbf{(EN)}} \textit{Obsessive-Compulsive Disorder Clinical Manifestations Prevalent Population Anxiety disorders are related to genetic factors, personality traits, adverse events, and stress factors, particularly closely linked to the patient's personality traits. For instance, excessive perfectionism, indecisiveness, meticulousness, and stubbornness are traits that increase the risk of developing OCD...}\\  
    \midrule
    \rowcolor{gray!20}\textbf{DXYDisease} \\
    \midrule
    \begin{CJK}{UTF8}{gbsn}
        \textbf{Query:} 维生素 A 缺乏症者需要做哪些检查来诊断？
    \end{CJK}\\
    \textit{\textbf{(EN)}} \textit{What tests are needed to diagnose vitamin A deficiency?}\\
      \begin{CJK}{UTF8}{gbsn}
        \textbf{Document:} 最准确的就是血液学检查。抽血检查血清维生素 A 的水平，对于成人来说，如果在 1.05～3.15 μmol/L，那么就表明不存在维生素 A 缺乏。 如果低于参考范围下限，那就是维生素 A 缺乏了。 ...
    \end{CJK}\\
    \textit{\textbf{(EN)}} \textit{The most accurate test is a hematological examination. A blood test to check the serum vitamin A levels is conducted. For adults, if the levels are between 1.05 and 3.15 umol/L, it indicates that there is no vitamin A deficiency. If the levels are below the lower limit of the reference range, it indicates vitamin A deficiency....}\\ 
    \midrule
    \rowcolor{gray!20}\textbf{MedicalRetrieval} \\
    \midrule
    \begin{CJK}{UTF8}{gbsn}
        \textbf{Query:} 一般宝宝的肚脐眼要多久愈合？
    \end{CJK}\\
    \textit{\textbf{(EN)}} \textit{How long does it take for a baby's belly button to heal?}\\
      \begin{CJK}{UTF8}{gbsn}
        \textbf{Document:} 你好，宝宝的肚脐一般是1-2周左右会好的，时间长的也有一个月的，不过这个时候可能会有脐茸了。
    \end{CJK}\\
    \textit{\textbf{(EN)}} \textit{Hello, a baby's belly button generally heals in about 1 to 2 weeks, although it may take up to a month in some cases. During this time, there might also be umbilical granuloma.}\\ 

    \midrule
    \rowcolor{gray!20}\textbf{CmedqaRetrieval} \\
    \midrule
    \begin{CJK}{UTF8}{gbsn}
        \textbf{Query:} 甲状腺手术后多久可以干活？
    \end{CJK}\\
    \textit{\textbf{(EN)}} \textit{How long after thyroid surgery can one return to work?}\\
      \begin{CJK}{UTF8}{gbsn}
        \textbf{Document:} 皮肤的修复一般由两周左右就会不影响你的颈部活动了，至于皮下软组织以及肌肉组织的修复可能时间长一下，一般一个月后就不会有明显影响了，你就可以工作了。工作中注意不要劳累，调整好自己的情绪。
    \end{CJK}\\
    \textit{\textbf{(EN)}} \textit{The skin usually heals in about two weeks, and you should no longer have restrictions on neck movement. However, the repair of subcutaneous soft tissue and muscle tissue may take longer. Generally, after about a month, there should be no significant impact, and you can return to work. During work, be sure to avoid overexertion and manage your emotions well.}\\
    \bottomrule
    \end{tabular}
    \caption{Data example in CMIRB (I).}
    \label{tab:data_example_1}
\end{table*}

\begin{table*}[!t]
    \centering
    \fontsize{9pt}{11pt}\selectfont
    \begin{tabular}{p{0.98\linewidth}}
    \midrule
        \rowcolor{gray!20}\textbf{DXYConsult} \\
    \midrule
    \begin{CJK}{UTF8}{gbsn}
        \textbf{Query:} 症状及患病时长：感冒，鼻炎，失去嗅觉一周。就医及用药情况：未就医，自行服用泰诺。需要解答的问题：鼻炎，失去嗅觉怎么办
    \end{CJK}\\
    \textit{\textbf{(EN)}} \textit{Symptoms and Duration of Illness: Cold, rhinitis, loss of smell for one week. Medical Consultation and Medication: No medical consultation, self-medicated with Tylenol. Questions Needing Answers: What to do about rhinitis and loss of smell?}\\
      \begin{CJK}{UTF8}{gbsn}
        \textbf{Document:} 你好，如果近期有这种感冒的病史的话，就会导致出现嗅觉功能下降，建议在口服感冒药的技术上的话，用海盐水冲洗鼻腔，一天两次，鼻喷辅舒良或者内舒拿看看效果，如果分泌过多的话，可以口服桉柠蒎胶囊，每天三次每次一粒。
    \end{CJK}\\
    \textit{\textbf{(EN)}} \textit{Hello, if there has been a recent history of cold symptoms, this can lead to decreased olfactory function. It is recommended to use saline nasal irrigation twice a day while taking cold medicine. You may also try nasal sprays like Budesonide or Fluticasone to see if they help. If there is excessive secretion, you can take Eucalyptus and Menthol capsules, three times a day, one capsule each time.}\\  
\midrule
        \rowcolor{gray!20}\textbf{CovidRetrieval} \\
    \midrule
    \begin{CJK}{UTF8}{gbsn}
        \textbf{Query:} 如何对待因履行工作职责感染新冠肺炎的医务人员？
    \end{CJK}\\
    \textit{\textbf{(EN)}} \textit{How should healthcare workers who contract COVID-19 while fulfilling their duties be treated?}\\
      \begin{CJK}{UTF8}{gbsn}
        \textbf{Document:} ...为进一步加强疫情防控期间医务人员防护工作，切实保障医务人员身心健康，现将有关要求通知如下：一、高度重视医务人员防护工作做好医务人员防护工作，是预防和减少医务人员感染的关键举措，...
    \end{CJK}\\
    \textit{\textbf{(EN)}} \textit{...To further enhance the protection of healthcare workers during the pandemic and ensure their physical and mental well-being, the following requirements are hereby notified: Pay great attention to the protection of healthcare workers
Ensuring proper protection for healthcare workers is a key measure in preventing and reducing infections among them, ...}\\  
    \midrule
    \rowcolor{gray!20}\textbf{IIYiPost} \\
    \midrule
    \begin{CJK}{UTF8}{gbsn}
        \textbf{Query:} 病例讨论：静脉输入阿昔洛韦2天，出现腰痛、尿少
    \end{CJK}\\
    \textit{\textbf{(EN)}} \textit{Case Discussion: Two days of intravenous acyclovir, followed by lower back pain and reduced urine output}\\
      \begin{CJK}{UTF8}{gbsn}
        \textbf{Document:} 1.病例资料,患者，男，31岁。因静脉输入阿昔洛韦2天，出现腰痛、尿少伴恶心、呕吐6天入院。患者8天前因受凉感冒，出现咳嗽、发热（最高体温38.6℃），无明显咳痰，院外静脉给予NS500ml+青霉素钠盐800万U，vd，1次/日,...
    \end{CJK}\\
    \textit{\textbf{(EN)}} \textit{Case Data, Patient: Male, 31 years old. The patient was admitted after experiencing lower back pain and reduced urine output, accompanied by nausea and vomiting for six days following two days of intravenous acyclovir administration. Eight days prior, the patient had caught a cold due to exposure, presenting with a cough and fever (highest temperature of 38.6°C), without significant sputum production. He received intravenous administration of ...}\\ 
    \midrule
    \rowcolor{gray!20}\textbf{CSLCite} \\
    \midrule
    \begin{CJK}{UTF8}{gbsn}
        \textbf{Query:} 微球在组织工程中的应用
    \end{CJK}\\
    \textit{\textbf{(EN)}} \textit{Application of Microspheres in Tissue Engineering}\\
      \begin{CJK}{UTF8}{gbsn}
        \textbf{Document:} 背景:骨组织工程骨构建中如何使生长因子持续高效发挥作用是影响成骨速度和质量的关键,现多以各种材料的微球或支架作为缓释载体,但缓释作用有待提高.目的:实验拟制备壳聚糖微球,然后复合到纳米羟基磷灰石/聚乳酸羟基乙酸支架上...
    \end{CJK}\\
    \textit{\textbf{(EN)}} \textit{Background: In bone tissue engineering, maintaining the sustained and efficient activity of growth factors is key to influencing the speed and quality of bone formation. Currently, microspheres or scaffolds made from various materials are commonly used as sustained-release carriers, but the release efficiency needs improvement. Objective: This experiment aims to prepare chitosan microspheres and incorporate them into a nano-hydroxyapatite/polylactic-co-glycolic acid (nHA/PLGA) scaffold, ...}\\ 

    \midrule
    \rowcolor{gray!20}\textbf{CSLRel} \\
    \midrule
    \begin{CJK}{UTF8}{gbsn}
        \textbf{Query:} 高血压病的辨治及预防 高血压病可归属中医学"眩晕"、"头痛"等范畴,其起病隐匿,不易引起患者的充分重视,中后期可致心脑血管疾病、肾损害...
    \end{CJK}\\
    \textit{\textbf{(EN)}} \textit{Differentiation and Treatment of Hypertension and Its Prevention Hypertension can be categorized under the terms "dizziness" and "headache" in traditional Chinese medicine (TCM). Its onset is insidious, often not receiving enough attention from patients, ...}\\
      \begin{CJK}{UTF8}{gbsn}
        \textbf{Document:} 辨证施治高血压 高血压病是现代医学病名,在中医归属眩晕病范畴,中医认为高血压与风、火、痰、虚有关.高血压的界定根据世界卫生组织(WHO)的标准,成人在休息状态下,收缩压持续高于140毫米汞柱...
    \end{CJK}\\
    \textit{\textbf{(EN)}} \textit{TCM Syndrome Differentiation and Treatment of Hypertension
Hypertension is a modern medical term, categorized under dizziness in TCM. TCM holds that hypertension is related to wind, fire, phlegm, and deficiency ...}\\
    \bottomrule
    \end{tabular}
    \caption{Data example in CMIRB (II).}
    \label{tab:data_example_2}
\end{table*}

\end{document}